# Characterizing Classes of Potential Outliers through Traffic Data Set Data Signature 2D nMDS Projection[*]

Erlo Robert F. Oquendo, Jhoirene B. Clemente, Jasmine A. Malinao, Henry N. Adorna
Department of Computer Science (Algorithm and Complexity Laboratory), University of the Philippines
Velasquez Ave., Diliman, Quezon City 1101
Email: {efoquendo, jbclemente}@up.edu.ph, {jamalinao, hnadorna}@dcs.upd.edu.ph

## ABSTRACT

This paper presents a formal method for characterizing the potential outliers from the data signature projection of traffic data set using Non-Metric Multidimensional Scaling (nMDS) visualization. Previous work had only relied on visual inspection and the subjective nature of this technique may derive false and invalid potential outliers. The identification of correct potential outliers had already been an open problem proposed in literature. This is due to the fact that they pinpoint areas and time frames where traffic incidents/accidents occur along the North Luzon Expressway (NLEX) in Luzon.

In this paper, potential outliers are classified into (1) absolute potential outliers; (2) valid potential outliers; and (3) ambiguous potential outliers through the use of confidence bands and confidence ellipse. A method is also described to validate cluster membership of identified ambiguous potential outliers.

Using the 2006 NLEX Balintawak Northbound (BLK-NB) data set, we were able to identify two absolute potential outliers, nine valid potential outliers, and five ambiguous potential outliers.

In a literature where Vector Fusion was used, 10 potential outliers were identified. Given the results for the nMDS visualization using the confidence bands and confidence ellipses, all of these 10 potential outliers were also found and 8 new potential outliers were also found.

## 1. INTRODUCTION

Data signatures [1, 2, 3] had long been used in literature to represent large data sets in smaller and more compact forms. Data signatures were initially used for clustering purposes, such as finding groups of related medical documents [1], finding traffic flow regularities and irregularities, and extracting expected, and/or sometimes unknown outliers from time series traffic volume data set [2]. As further intra-cluster and inter-cluster analysis were pursued, the need to qualitatively measure the goodness of visual representation arose. In our previous work [3], this measure was formulated as a set of 5 criterions. One such criterion, i.e. the easy detection of *potential outliers*, proved to be a very difficult and biased task. As defined in the paper, potential outliers are those points found *near* or at the periphery of a region occupied by a cluster in the $2-$dimensional visualization. These points may not have been categorized as outright outliers by the implemented clustering algorithm, yet, they show unique deviations at certain time intervals compared to their co-members of their clusters. These points had been shown to provide areas and time frames of interest to domain experts as they show occurrences of traffic incidents. However, note that *nearness* to a periphery of a cluster's region has not yet been quantitatively described, thus, the difficulty. The issue of providing definite quantitative descriptions or definitions has yet to done.

Nonetheless, by realizing the data signatures' potential in obtaining more refined cluster models than time-domain based clustering for a given time series traffic data set, we conducted further studies in [4] to check the effectiveness of using data signatures in producing its $2-$dimensional Vector Fusion-based visualizations. We were able to show that, indeed, instead of using $n$ dimensions, where $n$ is the number of dimensions of the time- and frequency-domain, we use only $7-$dimensional (frequency-domain) data signatures. By doing so, we are able to capture more information and a better visual representation of a time-domain $n$-dimensional data set. In addition to this result, we were able to provide an algorithm quantitatively measuring the consistency of a $2D$ data signature-based visualization with the original $n$-dimensional frequency-domain representation of the given data set. Venturing into the use of other visualization tools and implementing this algorithm on the visualizations produced, we were able to show in [5] that a non-Metric Multidimensional Scaling (nMDS) visualizations achieve better consistency against Vector Fusion-based visualizations.

In this paper, we address the formalization of the aforementioned potential outliers. We present a method pinpointing these points using confidence bands and confidence ellipses derived from fitted curves for each cluster in the data set. Furthermore, by using these tools, we further refine and provide better categorization of potential outliers, namely, a) absolute potential outliers, b) ambiguous potential outliers, and c) valid potential outliers.

Section 2 provides the core definitions and notations used in the paper. We mention the data set, pre-processing methods and data signature generation of the dataset. We discuss the computations done for non-Metric Multidimensional Scaling and how confidence bands and confidence ellipses are derived from 2D nMDS projections of the data signatures for potential outlier detection. Section 3 explains how outliers are characterized based on the methods presented in the previous section and how cluster membership may be validated when ambiguity exists. Section 4 discusses the results of the characterization of outliers as well as the validation of cluster membership in the traffic data set. Finally, we provide the conclusions in Section 5.

[*] This work is partially funded by the Engineering Research and Development for Technology "Information Visualization via Data Signatures" 2009-2011

## 2. DEFINITIONS AND BASIC NOTATIONS

### 2.1 Data Set
The data sets used in this work have the following characteristics: time-series, periodic and multidimensional. In particular, we used records in the 2006 NLEX Balintawak Northbound (BLK-NB) data set provided by the National Center of Transportation Studies (NCTS). Data set contains the hourly traffic volume of lanes along Balintawak Northbound (BLK-NB) of North Luzon Express way (NLEX). An automatic detector was embedded on each lane of the corridor which captures the mean spot traffic volume per hour. The record from NCTS provided the hourly traffic volume and speed for the whole year of 2006, giving a total of 52 weeks x 168 time points.

The data set were filtered so that weeks with unreasonable value for traffic volume were eliminated. The corresponding set of speeds for each week was also taken into account; values that don't satisfy the defined thresholds were also eliminated. The result is a $42 \times 168$ data matrix for the analysis.

### 2.2 Data Signature
A *data signature*, as defined in [1] is a mathematical data vector that captures the essence of a large data set in a small fraction of its original size. These signatures allow us to conduct analysis in a higher level of abstraction and yet still reflect the intended results as if we are using the original data.

In this paper, we use Power Spectrum-based data signatures for time series traffic data representation for visualization purposes. An optimal set of Power Spectrum values were determined in [4] by use of the characteristics of the frequency-domain representation and the quantitative goodness-of-visualization measure defined in the same paper. Note that these visualizations are already produced using the non-Metric Multi-dimensional Scaling(nMDS) algorithm since we have already shown in [5] that these visualizations better represent the original data set compared with Vector Fusion-based visualizations. We take as data signature of each week from the 2006 BLK-NB traffic volume data set the offset, and the $7th, 14th, 21st, 28th, 35th, 42nd$ harmonics from the 168 Power Spectrum components.

### 2.2 Power Spectrum
Fourier analysis is one well-known methodology for signal processing. By treating each *n*-dimensional weekly partitions in the NLEX BLK-NB time-series traffic data set as discrete signals, we would be able to transform each partition into their more compact, frequency-domain representation. Fourier descriptors, can then be scrutinized to obtain an optimal selection of frequency-domain components to represent each signal in lower dimensions. One such descriptor, that is Power Spectrum, rely on the fact that any signal can be decomposed into a series of frequency components via Discrete Fourier Transforms, as shown below,

$$\theta(t) = \mu_0 + \sum_{k=1}^{n-1} (a_k \cos \frac{2\pi k}{n} - i b_k \sin \frac{2\pi k}{n}),$$

where $\mu_0$ is the component referred to as the *offset* of the signal translated from the horizontal axis. Using DFT, a vector of real numbers can produce a vector $F$ of frequency components of the same length, where $F = (a_0 \pm b_0 i, a_1 \pm b_1 i, a_2 \pm b_2 i, \ldots, a_{n-1} \pm b_{n-1} i)$, $\mu_0 = a_0 \pm b_0 i$ and $i^2 = -1$. For the resulting $nD$ vector, we produce distinct values for $a_0 \pm b_0 i, a_1 \pm b_1 i, a_2 \pm b_2 i, \ldots, a_{n/2} \pm b_{n/2} i$ and the succeeding values are their complex conjugates. If $n$ is even, the $(n/2 + 1)^{th}$ component will be real.

Power Spectrum is the distribution of power values as a function frequency. For every frequency component, power can be measured by summing the squares of the coefficients of the corresponding $sine/cosine$ pair and then getting its square root. The Power Spectrum $A_k$ of the signal $k = 1, 2, \ldots, n-1$ is given by

$$A_k = \sqrt{a_k^2 + b_k^2}.$$

### 2.3 Non-Metric Multidimensional Scaling
The data signature representation of the data set was visualized through nMDS [3]. It showed that the projection conforms to the defined criteria in [1].

Non-metric multidimensional scaling or nMDS is a dimensionality reduction technique that transforms an $n \times m$ data matrix to a $2 - dimensional$ Euclidean space $E$, such that the properties of the $n$ objects in $O$ are, as much as possible, captured in the lower dimensional space[6]. The algorithm takes the dissimilarity matrix $D$ obtained through pairwise computation of the distances of each point in $O$ as input. Each object in $D$ is computed as

$$D_{ij} = (x_i - x_j)^T (x_i - x_j)$$

where $x_i$ and $x_j$ are elements of $O$ and $D_{ij} = D_{ji}$. From the dissimilarity matrix $D$, an inner product $B$ is defined such that $B = X^T X$ where $X$ consists of the coordinates of the objects in the Euclidian space. Spectral decomposition of B allows for the computation of $X$ by obtaining the set of eigenvalues and eigenvectors of $B$. Since $B$ is a positive-definite matrix with rank $p$, $p$ non-zero eigenvalues and $n - p$ zero eigenvalues from $B$ may be obtained. This property of the matrix $B$ allows for $X$ with a dimension of $n \times p$ to be calculated. The resulting matrix $X$ is projected in a $p$ euclidean space.

### 2.4 Confidence Intervals

#### 2.4.1 Confidence Band
A confidence interval with a confidence coefficient $(1 - \alpha)$, $0 \leq \alpha \leq 1$, is random interval whose endpoints are statistics called confidence limits. A $100(1 - \alpha)\%$ confidence interval is given a $100(1 - \alpha)\%$ confidence to contain the true value of the parameter estimated, e.g. mean $\mu$.

The idea of confidence intervals may be extended to curve estimation where the confidence limit for every value of $x$ on the curve is plotted along with the estimated curve. A confidence band encloses an area that one can be $100(1 - \alpha)\%$ certain contains the true curve. It gives a visual sense of how well the data define the best-fit curve. A confidence band is constructed as follows.

The best-fit curve is constructed. The confidence band is extended above and below the curve by

$$\sqrt{c}\sqrt{\frac{SS}{DF}}t_\alpha(DF)$$

where $c = G|x \times \Sigma \times G'|x$, $G|x$ is the gradient vector of the parameters at a particular value of x, $G'|x$ is the transposed gradient vector, $\Sigma$ is the variance-covariance matrix, SS is the sum of squares for the fit, DF is the degrees of freedom, and $t_\alpha(DF)$ is the Student's t critical value based on the confidence level α and the degrees of freedom $DF$.

### 2.4.2 Confidence Ellipse
Confidence ellipse is another plot related to the confidence band. It uses intervals for both $X$ and $Y$. The interval is projected horizontally and vertically respectively. The confidence ellipse is formed by the following equation

$$\bar{Z} \pm R \times I$$

where $\bar{Z}$ is the mean of either $X$ or $Y$, $R$ is the range of either $X$ or $Y$, and $I$ is the confidence level $1 - \alpha$. These form the minor and major axes of the ellipse. The ellipse is given a $100(1 - \alpha)\%$ confidence to contain the data points it bounds.

### 2.4.3 Potential Outliers
*Potential outliers,* as previously defined in [3] are points projected "near" or at the periphery of a region occupied by its cluster in the $2$−dimensional visualization. By studying the original time series data set more closely, it was found out in the same paper that these points are indicative of traffic incidents, i.e. prolonged very low or high traffic volume values, significant deviation of a week's behaviour as the first few data values were collected, or abrupt change and fluctuations of traffic volume values within a day of the week, etc. Thus, pinpointing these set of weeks is also of interest to domain experts for NLEX. Even when the correctness of identifying these set of weeks had been validated by domain experts, we still have not provided a better and more definite way of defining these points. Establishing the minimum Euclidean distance from a cluster's group of points or a point of reference, such as its center of mass, has not been determined so as to classify one member of a cluster as a potential outlier. We now propose a formal way of defining and identifying such type of points as seen in the succeeding subsections.

#### 2.4.3.1 Absolute Potential Outliers
An absolute potential outlier is a point lying outside the confidence band and confidence ellipse. This point is no longer bounded by the confidence ellipse and is not represented by fitted curve.

#### 2.4.3.2 Valid Potential Outliers
A valid potential outlier is a point lying outside the confidence ellipse but is still within the confidence band. This point is no longer bounded by the confidence ellipse but is still represented by fitted curve.

#### 2.4.3.3 Ambiguous Potential Outliers
An ambiguous potential outlier is a point that is bounded by two different confidence ellipses or two different confidence bands, or a point that is within the confidence ellipse but outside the confidence band. It is unclear as to which cluster should this point be identified with.

## 3. METHODOLOGY
The remaining set of weeks for the 2006 NLEX BLK-NB is used to obtain 42 data signatures for further traffic analysis. We adapt the same (optimal) cluster model we obtained in [3] and use it to color points in the data signature-based nMDS projection of the data set. Each of the $7 - dimensional$ data signatures is now represented as $2 - dimensional$ points in Euclidean space. Then, through this projection, we derive the best-fit polynomial curves representing the behavior of the points within each cluster in the data set. Curve fit is determined using the coefficient of determination $R^2$. By using a 95% confidence level, we determined the (95%) confidence bands and (95%) confidence ellipses for all surrounding regions occupied by the clusters. Guided by the definitions of potential outliers in Section 1, we pinpoint the set of Valid Potential Outliers, Ambiguous Potential Outliers, and Absolute Potential Outliers. Finally, we check the consistency of the obtained potential outliers based on the nMDS visualization from the set of potential outliers identified in our previous work where Vector Fusion was used [3,4].

### 3.1 Characterizing Outliers
The confidence bands and confidence ellipse are used in the intra-cluster and inter-cluster analyses. The bands and the ellipses are used to determine whether a point is an outlier with respect to the cluster it belongs to or not. The fitted curves, being a representation of the clusters, are used to check if a point is closer to a different cluster than it is presently identified with.

One can give a confidence level that the band or the ellipse contains the true members of the clusters. Hence, both act as thresholds for cluster membership. Points lying outside the confidence bands of the cluster are said to be outliers with respect to the cluster they are identified with. The interpretation of the outliers is done with respect to the *y-axis*.

The confidence ellipse gives more information about the outliers. Points lying outside the ellipses of the clusters are also identified as outliers but the interpretation is done with respect to the *x-axis* and the *y-axis*. Both the confidence band and the confidence ellipse are used to characterize the potential outlier as being an absolute potential outlier, valid potential outlier, or ambiguous potential outlier.

The ambiguity of the cluster membership of ambiguous potential outliers will be removed since a point should belong to only one cluster. The distances of the point to the curves fitted on the different clusters will be determined. The point will be identified with the cluster it is most proximate with. That is, a point will be identified with a cluster if the distance of the point to the curve of that cluster is the minimum distance from among the distances of the point to the curves of the different clusters. If the curve is a polynomial curve of degree one, then the distance will simply be the perpendicular distance of the point to the line. Otherwise, the shortest distance from the point to the curve may be found by taking the derivative of the function at that point and set it to zero.

## 4. RESULTS

Potential outliers per cluster are classified using the confidence bands and ellipses constructed from the curves fitted per cluster. No curves were fitted for clusters 2, 5, 6, and 7 because there are not enough data on those clusters to construct the curves. Table 1 shows the fitted curves per cluster.

**Table 1. Curves fitted per cluster**

| Cluster | Curve |
|---|---|
| 0 | $y = -0.234 + 1.462x - 2.244x^2 + 1.117x^3$ |
| 1 | $y = 0.052 + 0.028x - 0.355x^2$ |
| 3 | $y = 0.149 - 0.092x$ |
| 4 | $y = -1.612 - 3.190x - 1.426x^2$ |

95% confidence bands are constructed for every fitted curve. 95% confidence ellipses are also constructed based on the points on different clusters. Figure 1 shows the confidence bands and ellipses for the different clusters. Week 1 is an absolute outlier within cluster 0. It lies outside the band and the ellipse. Week 1 may belong to or form a separate cluster from cluster 0. Week 32 is a valid potential outlier within cluster 0. It lies within the band but outside the ellipse. Weeks 27 and 8 are ambiguous outliers within cluster 0. They lie within the ellipse but outside the band.

**Figure 1. Confidence bands and ellipses per cluster**

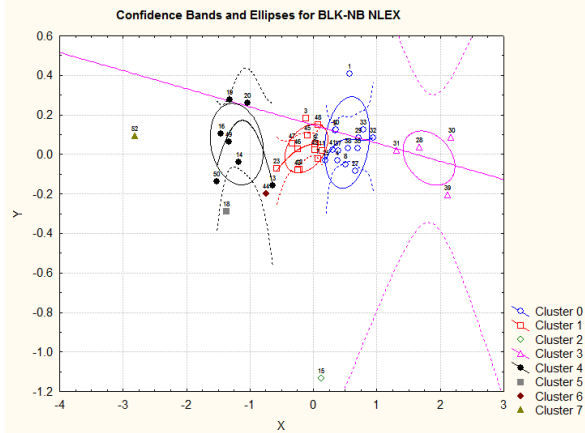

A summary of the classification of potential outliers is given in Table 2. No point lies within two different ellipses. Weeks 15, 18, 44, and 55 are known outliers in previous studies, forming separate clusters 2, 5, 6, and 7 respectively. The ambiguous potentials outliers identified here are those that lie outside the confidence band but within the confidence ellipse.

**Table 2. Summary of Classification of Potential Outliers**

| Cluster | Absolute Potential Outliers | Valid Potential Outliers | Ambiguous Potential Outliers |
|---|---|---|---|
| 0 | Wk1 | Wk32 | Wk27, Wk8 |
| 1 | Wk3 | Wk23 | Wk48, Wk12, Wk43 |
| 3 | | Wk30, Wk31, Wk39 | |
| 4 | | Wk19, Wk20, Wk50, Wk13 | |

## 5. CONCLUSION AND RECOMMENDATION

Confidence bands and confidence ellipse were used to characterize potential outliers. Potential outliers are characterized as absolute, valid, or ambiguous potential outliers based on whether they lie or not within the confidence band and confidence ellipse. Points that lie outside the confidence band and ellipse are identified as absolute potential outliers. Points that lie within the confidence band but outside the confidence ellipse are indentified as valid potentials outliers. Finally, points that either lie within two confidence bands or confidence ellipse or lie within the confidence ellipse but outside the confidence band are identified as ambiguous potential outliers. Ambiguous potential outliers may be assigned to a cluster that it is most proximate with. To determine proximity of points to clusters, the distance of the points to the curves fitted per cluster is used. For polynomial curves of degree one, this distance will simply be the perpendicular distance of the point to the curve. For polynomial curves of higher degree, this distance is determined by solving for the slope of the tangent line to that curve that passes through the given point. The distance will be the length of the line that connects the point and the tangential point of the curve.

Since no ambiguous potential outliers are identified in the data, the method described for validating cluster membership for ambiguous potential outliers may be tried on a different data set. Only polynomial curves were fitted. Future work may also fit non-polynomial curves.

Previous studies using Vector Fusion have found 10 potential outliers. All of these were also found in the new nMDS visualization using confidence bands and confidence ellipses along with 8 new potential outliers.

## 6. ACKNOWLEDGMENTS

The authors would like to thank Engineering Research and Development for Technology (ERDT) for funding this paper under the project entitled "Information Visualization via Data Signatures".

Erlo Robert F. Oquendo would like to thank the University of the Philippines Visayas for his Fellowship Grant. Jhoirene Clemente would like to thank ERDT for funding her graduate studies in UP Diliman.

## 7. REFERENCES

[1] P.C. Wong, H. Foote, R. Leung, D. Adams and J. Thomas, Data Signatures and Visualization of Scientific Data Sets. Pacific Northwest National Laboratory, USA, IEEE 2000.

[2] J. A. Malinao, R.A. B. Juayong, F.J. O. Corpuz, J.M. C. Yap, H. N. Adorna. Data Signatures for Traffic Data Analysis. Presented in the National Conference for Information Technology Education 2009, October 2009.

[3] J. A. Malinao, R.A. B. Juayong, J. G. Becerral, K.R. C. Cabreros, K.M. B. Remaneses, J. G. Khaw, D. F. Wuysang, F.J. O. Corpuz, N.H. S. Hernandez, J.M. C. Yap, and H. N. Adorna, "Patterns and Outlier Analysis of Traffic Flow using Data Signatures via BC Method and Vector Fusion Visualization", To appear in Proc. of the 3rd International

Conference on Human-centric Computing (HumanCom-10), 2010.

[4] J. A. Malinao, R.A. B. Juayong, and H. N. Adorna, Improving Data Signature Construction through Vector Fusion Quantitative Analysis, To appear in the Proc. of the Conference of Visualization and Data Analysis 2011, 2010.

[5] J. B. Clemente, R. U. Tadlas, R.A. B. Juayong, J. A. Malinao, and H. N. Adorna, Visualization of Data Signatures using Non-Metric Multidimensional Scaling for Time Series Traffic Data Analysis, To appear in the Proceedings of the 2010 International Congress and Computer Applications and Computational Science (CACS 2010), 2010.

[6] T.F. Cox and M.A. Cox, Multidimensional Scaling, 1994

# Appendix

## Confidence Bands for the different Clusters

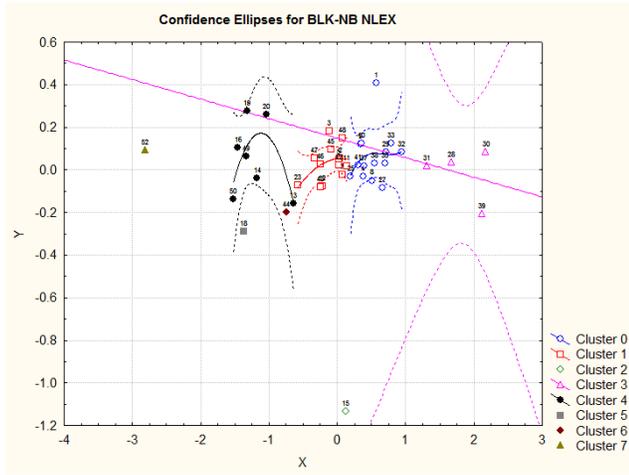

## Confidence Ellipses for the different Clusters

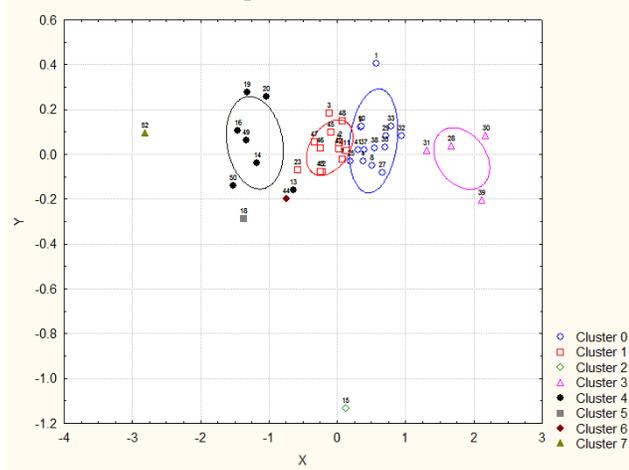